\documentstyle[12pt]{article}

\newcommand{\beq}{\begin{equation}}
\newcommand{\eeq}{\end{equation}}
\newcommand{\beqa}{\begin{eqnarray}}
\newcommand{\eeqa}{\end{eqnarray}}
\newcommand{\pa}{\partial}

\newenvironment{eqabc}%
{\setcounter{enumi}{\value{equation}}%
\addtocounter{enumi}{1}%
\setcounter{equation}{0}%
\begin{eqnarray}}%
{\end{eqnarray}\setcounter{equation}{\value{enumi}}}%

\newcommand{\ee}{\mbox{e}}
\newcommand{\dd}{\mbox{d}}
\newcommand{\sgn}{\mbox{\rm sgn}}
\newcommand{\Pf}{\mbox{\rm Pf}}

\title{Orthogonal and Symplectic Matrix Integrals and Coupled KP Hierarchy}
\author{Saburo Kakei\footnote{
Present Address: Department of Mathematical Sciences, Waseda University, 
Ohkubo 3-4-1, Shinjuku-ku, Tokyo 161-8555, Japan}\\[5mm]
{\it Department of Mathematical Sciences, University of Tokyo,}\\
{\it Komaba 3-8-1, Meguro-ku, Tokyo 153-8902, Japan}}
\date{}

\begin{document}
\maketitle

\begin{abstract}
Orthogonal and symplectic matrix integrals are investigated. 
It is shown that the matrix integrals can be considered as a 
$\tau$-function of the coupled KP hierarchy, 
whose solution can be expressed in terms of pfaffians.
\end{abstract}
\bigskip
\bigskip

Over the past decade, the intimate relationships between matrix 
integrals and nonlinear integrable systems have been clarified, 
particularly in the context of string theory. 
In such cases, nonperturbative properties of physical quantities 
can be evaluated by the use of the integrable structures of the models. 
(For review, see refs. 1-4.)

Here, we consider a matrix integral over an ensemble of 
Hermitian matrices 
\beq
Z_N^{(\beta)}\{t\} = \frac{1}{N!}
\int_{-\infty}^{\infty}\cdots\int_{-\infty}^{\infty} 
\prod_{1\leq j<k \leq N} |x_j-x_k|^{\beta} 
\exp\left[ \sum_{i=1}^{N} \eta(x_i,t) \right]
\dd x_1 \cdots \dd x_N, \label{Herm}
\eeq
where $\beta=1$, $2$ or $4$, for which the ensemble is, respectively, 
orthogonal, unitary or symplectic and 
$\eta(x,t) = \sum_{m=1}^{\infty}x^m t_m$. 
An interesting observation is that, when $\beta=2$, the quantity 
$Z_N^{(\beta=2)}\{t\}$ is a $\tau$-function of the Kadomtsev-Petviashvili 
(KP) hierarchy \cite{Mor,MMM,Mulase1,Mulase2}. 
In fact, the integral (\ref{Herm}) can be identified as a continuum 
limit of the multisoliton solution of the hierarchy.
This fact may be easily observed if we rewrite $Z_N^{(\beta=2)}$ in the 
following form: 
\beq
Z_N^{(\beta=2)}
= N!\det\left[\int_{-\infty}^{\infty}
x^{j+k-2}\ee^{\eta(x,t)}
\dd x\right]_{j,k=1,\cdots,N}. 
\eeq
It may be also clear that the quantity $Z_N^{(\beta=2)}$ 
satisfies the bilinear Toda equation 
\beq
Z_N^{(\beta=2)}\cdot\pa_{t_1}^2 Z_N^{(\beta=2)}
- \left(\pa_{t_1}Z_N^{(\beta=2)}\right)^2
+ Z_{N+1}^{(\beta=2)}\cdot Z_{N-1}^{(\beta=2)} = 0. 
\label{biToda}
\eeq

Here, we give other examples of such kinds of relationship. 
The orthogonal case $\beta=1$ and the symplectic case 
$\beta=4$ lead to $\tau$-functions of the coupled KP hierarchy 
proposed by Hirota and Ohta \cite{HO}. 
The coupled KP hierarchy is an extension of the KP hierarchy. 
The first few bilinear equations of the hierarchy are 
\begin{eqabc}
&&(D_1^4-4D_1D_3+3D_2^2)\tau\cdot\tau=24\bar{\sigma}\sigma,
\label{bi1}\\
&&(D_1^3D_2+2D_2D_3-3D_1D_4)\tau\cdot\tau=12D_1\bar{\sigma}\cdot\sigma,
\label{bi2}\\
&&(D_1^3+2D_3+3D_1D_2)\sigma\cdot\tau=0,\label{bi3}\\
&&(D_1^4-4D_1D_3-3D_2^2-6D_4)\sigma\cdot\tau=0,\label{bi4}\\
&&(D_1^3+2D_3-3D_1D_2)\bar{\sigma}\cdot\tau=0,\label{bi5}\\
&&(D_1^4-4D_1D_3-3D_2^2+6D_4)\bar{\sigma}\cdot\tau=0, \label{bi6}
\end{eqabc}%
where the Hirota bilinear operators are defined as
\[
D_m^k D_n^l f\cdot g = \left. 
\left(\frac{\pa}{\pa t_m}-\frac{\pa}{\pa t'_m}\right)^k
\left(\frac{\pa}{\pa t_n}-\frac{\pa}{\pa t'_n}\right)^l
f(t_1,t_2,\ldots)g(t'_1,t'_2,\ldots)\right|_{t'=t}. 
\]
If we set 
\[
u = 2\frac{\pa^2}{\pa t_1^2}\log \tau,\qquad
v = \sigma/\tau,\qquad
\bar{v} = \bar{\sigma}/\tau,
\]
we obtain the coupled KP equation 
from the bilinear equations (\ref{bi1}), (\ref{bi3}), (\ref{bi5}) as 
\begin{eqabc}
&& \frac{\pa}{\pa t_1}\left(4\frac{\pa u}{\pa t_3}
  -6u\frac{\pa u}{\pa t_1}-\frac{\pa^3 u}{\pa t_1^3}\right) 
  -3\frac{\pa^2 u}{\pa t_2^2} +24\frac{\pa^2}{\pa t_1^2}(v\bar{v})=0,\\
&& 2\frac{\pa v}{\pa t_3}+3u\frac{\pa v}{\pa t_1}+\frac{\pa^3 v}{\pa t_1^3}
  +3\left( \frac{\pa^2 v}{\pa t_1\pa t_2} + 
  v \int^{t_1} \frac{\pa u}{\pa t_2}\dd t_1\right) = 0,\\
&& 2\frac{\pa \bar{v}}{\pa t_3}+3u\frac{\pa \bar{v}}{\pa t_1}
  +\frac{\pa^3 \bar{v}}{\pa t_1^3} 
  +3\left( \frac{\pa^2 \bar{v}}{\pa t_1\pa t_2} + 
  \bar{v}\int^{t_1} \frac{\pa u}{\pa t_2}\dd t_1\right) = 0.
\end{eqabc}%
We note that the coupled KP hierarchy reduces to the ordinary KP 
hierarchy by setting $\sigma=0$ or $\bar{\sigma}=0$. In that case, 
the coupled KP equations (3) reduce to the KP equation. 

While the solutions of the KP hierarchy are expressed by
determinants, the solutions of the coupled KP hierarchy 
are expressed in terms of pfaffians \cite{HO}. 
We denote a pfaffian associated with the $2N\times2N$ antisymmetric 
matrix $A_{2N}=[a_{ij}]_{1\leq i,j\leq 2N}$ as
\[
\Pf[A_{2N}] = \mathop{{\sum}'}
\sgn\left(\begin{array}{cccc}
1 & 2 & \cdots & 2n\\
j_1 & j_2 & \cdots & j_{2n}
\end{array}\right) 
a_{j_1j_2}a_{j_3j_4}\cdots a_{j_{2n-1}j_{2n}} , 
\]
where the summation is taken over all $(j_1,\ldots,j_{2n})$ such that 
$j_1<j_3<\cdots<j_{2n-1}$ and $j_1<j_2$, $\ldots$, $j_{2n-1}<j_{2n}$. 
Using this notation, solutions of the bilinear equations of the 
coupled KP hierarchy can be constructed as
\begin{eqnarray}
\tau = \mbox{\rm Pf}[A_{2N}],\qquad
\sigma = \mbox{\rm Pf}[A_{2N-2}],\qquad
\bar{\sigma} = \mbox{\rm Pf}[A_{2N+2}], 
\label{tau}
\end{eqnarray}
where the elements $a_{l,m}$ satisfy the differential rules
\beq
\frac{\pa}{\pa t_n}a_{l,m} = a_{l+n,m} + a_{l,m+n},
\label{difrel}
\eeq
for $n=1,2,\ldots$. 
An example of the elements $a_{l,m}$ which satisfy eq. (\ref{difrel}) is 
\beq
a_{l,m} = \sum_{k}c_k (p_k^{l-1} q_k^{m-1} - p_k^{m-1} q_k^{l-1})
\exp\left[ \eta(p_k,t)+\eta(q_k,t) \right]. 
\label{element1}
\eeq
This $\tau$-function corresponds to a multisoliton solution. 
If $\tau$, $\sigma$ and $\bar{\sigma}$ are of the form in 
eq. (\ref{tau}), 
the bilinear equations (2a-f) reduce to algebraic 
identities of pfaffians \cite{HO}.

To obtain the relationship between the matrix integral and the coupled 
KP hierarchy, we use the equations attributable to de Bruijn \cite{Bruijn}: 
\beqa
\lefteqn{\stackrel{\displaystyle{\int\cdots\int}}{
\scriptstyle{x_1\leq\cdots\leq x_N}}
\raisebox{2.5ex}{$
\det\left[\phi_j(x_k)\right]_{j,k=1,\cdots,N}\,
\dd x_1\cdots \dd x_N$}}
\qquad\qquad \nonumber\\
&=& {\rm Pf}\left[\int\int{\rm sgn}(y-x)\,
\phi_j(x)\,\phi_k(y)\,\dd y\,\dd x\right]_{j,k=1,\cdots,N}
\nonumber\\
&=& 
{\rm Pf}\left[\,
\raisebox{-2.7ex}{$\displaystyle 
\stackrel{\displaystyle{\int\int}}{\scriptstyle{x<y}}$}
\left(\phi_j(x)\,\phi_k(y)- \phi_j(y)\,\phi_k(x)\right)
\,\dd y\,\dd x\right]_{j,k=1,\cdots,N}, 
\label{1det}\\
\lefteqn{
\int\cdots\int\,\det\left[\phi_j(x_k),\ \ \psi_j(x_k)\right]_{j=1,\cdots,2N,\
k=1,\cdots,N}\,\dd x_1\cdots \dd x_N}
\qquad\qquad \nonumber\\
&=& N!\;{\rm Pf}
\left[\int(\phi_j(x)\,\psi_k(x)-\phi_k(x)\,\psi_j(x))\,\dd x
\right]_{j,k=1,\cdots,2N}.
\label{4det}
\eeqa
We note that Tracy and Widom utilized these formulas to obtain the 
Fredholm determinant expressions of similar matrix integrals \cite{TW}. 
Using these formulas, 
we first consider the orthogonal matrix model.
Applying eq. (\ref{1det}) and the Vandermonde determinant
\[
\prod_{j>k}(x_j-x_k)=\det\left[x_k^{j-1}\right]_{j,k=1,\ldots,N}, 
\]
we can rewrite $Z_N^{(\beta=1)}\{t\}$ in terms of a pfaffian as 
\beqa
Z_N^{(\beta=1)}\{t\} &=&
\raisebox{-2.7ex}{$
\stackrel{\displaystyle{\int\cdots\int}}{
\scriptstyle{x_1\leq\cdots\leq x_N}}$}
\prod_{1\leq k<j \leq N} (x_j-x_k)
\exp\left[ \sum_{i=1}^{N}\eta(x_i,t) \right]
\dd x_1 \cdots \dd x_N\nonumber\\
&=& 
{\rm Pf}\left[\,
\raisebox{-2.7ex}{$\displaystyle 
\stackrel{\displaystyle{\int\int}}{\scriptstyle{x<y}}$}
(x^{j-1} y^{k-1} - y^{j-1} x^{k-1})
\exp\left[ \eta(x,t)+\eta(y,t) \right] \dd x\dd y
\right]_{j,k=1,\cdots,N}.
\label{Orth}
\eeqa%
Because the elements of eq. (\ref{Orth}) may be considered as a continuum 
limit of the multisoliton solution given by eq. (\ref{element1}), 
they satisfy the differential rules given by eq. (\ref{difrel}). 
Hence, we conclude that $Z_N^{(\beta=1)}\{t\}$ is a 
$\tau$-function of the coupled KP hierarchy, i.e., 
$Z_N^{(\beta=1)}\{t\}$ solves the bilinear equations (2a-f). 

We then consider the symplectic case. Toward this aim, we prepare another 
formula. Consider the twofold Vandermonde determinant
\beq
\det\left[x_k^{j-1},\; y_k^{j-1}\right]_{j=1,\ldots,2N,\ k=1,\ldots,N}
= \prod_{j>k}(x_j-x_k)\cdot\prod_{j>k}(y_j-y_k)\cdot
\prod_{j,k=1}^N(y_j-x_k). 
\eeq
Differentiating with respect to each $y_k$ and setting $y_k=x_k$, 
we obtain the following equation \cite{TW}: 
\beq
\prod_{j<k}(x_j-x_k)^4=\det\left[ x_k^j,\; (j-1)x_k^j \right]_
{j=1,\ldots,2N,\ k=1,\ldots,N}.
\label{pretty}
\eeq
Applying eqs. (\ref{4det}) and (\ref{pretty}), 
we can rewrite $Z_N^{(\beta=4)}\{t\}$ in terms of a pfaffian as 
\beqa
Z_N^{(\beta=4)}\{t\} &=& \frac{1}{N!}
\int_{-\infty}^{\infty}\cdots\int_{-\infty}^{\infty}
\prod_{1\leq j<k \leq N} (x_j-x_k)^4
\exp\left[ \sum_{i=1}^{N}\eta(x_i,t) \right]
\dd x_1 \cdots \dd x_N\nonumber\\
&=& 
{\rm Pf}\left[\,
\int_{-\infty}^{\infty}
(k-j)x^{k+j-3}
\exp\left[ 2\eta(x,t) \right] \dd x
\right]_{j,k=1,\cdots,N}.
\label{Symp}
\eeqa
Because the elements of eq. (\ref{Symp}) satisfy eq. (\ref{difrel}), 
$Z_N^{(\beta=4)}\{t\}$ is another example of a $\tau$-function of the 
coupled KP hierarchy. 

Lastly, we have demonstrated that the quantities 
\[
u^{(\beta)} = 2\frac{\pa^2}{\pa t_1^2}\log 
Z_N^{(\beta)}\{t\},\qquad
v^{(\beta)} = Z_{N-1}^{(\beta)}\{t\}/Z_N^{(\beta)}\{t\},\qquad
\bar{v}^{(\beta)} = Z_{N+1}^{(\beta)}\{t\}/Z_N^{(\beta)}\{t\},
\]
with $\beta=1,4$, 
satisfy the equations of the hierarchy such as the coupled KP 
equations (4). 
In the case $\beta=2$, nonperturbative properties of the 
``specific heat'' $u=2\pa_{t_1}^2\log Z_N^{(\beta=2)}\{t\}$ can be evaluated 
directly using the KP equation \cite{KKN}. 
However, the properties of the coupled KP hierarchy are not as well 
understood as those of the ordinary KP hierarchy.
For instance, Lax formulations and conserved quantities for the coupled 
hierarchy have not yet been obtained. 
Furthermore, although $Z_N^{(\beta=2)}$ satisfies the bilinear Toda 
equation (\ref{biToda}), 
the corresponding facts for the cases $\beta=1,4$ are not 
known to date. We hope to report them in a forthcoming article. 

\section*{Acknowledgments}
The author gratefully acknowledges Dr. Yasuhiro Ohta for discussions and 
helpful comments. 
This work was supported by J.S.P.S. Research Fellowships for 
Young Scientists.

\end{document}